\def\ifarxiv{\iftrue}     

\ifarxiv
\documentclass[12pt,a4paper]{article}

\usepackage{amsmath,amssymb}
\usepackage[nosort]{cite}
\usepackage[hyperref,parsep]{collect}

\makeatletter
\let\old@makecaption=\@makecaption
\def\@makecaption{\small\old@makecaption}
\makeatother

\else
\documentclass[pop,fleqn]{w-art}

\usepackage{times}

\chardef\bslash=`\\ 

\usepackage{amsfonts}
\usepackage[parsep]{collect}

\fi

\let\oldPhi=\Phi
\let\oldPsi=\Psi
\let\oldGamma=\Gamma
\let\oldDelta=\Delta
\let\oldSigma=\Sigma
\let\oldTheta=\Theta
\let\oldPi=\Pi
\renewcommand{\Phi}{\mathnormal{\oldPhi}}
\renewcommand{\Psi}{\mathnormal{\oldPsi}}
\renewcommand{\Gamma}{\mathnormal{\oldGamma}}
\renewcommand{\Sigma}{\mathnormal{\oldSigma}}
\renewcommand{\Delta}{\mathnormal{\oldDelta}}
\renewcommand{\Theta}{\mathnormal{\oldTheta}}
\renewcommand{\Pi}{\mathnormal{\oldPi}}

\newcommand{\ham}{\mathcal{H}}
\newcommand{\charge}{\mathcal{Q}}
\newcommand{\gen}[1]{\mathfrak{#1}}
\newcommand{\dil}{\gen{D}}
\newcommand{\len}{\mathcal{L}}

\newcommand{\opperm}{\mathcal{P}}
\newcommand{\superN}{\mathcal{N}}
\newcommand{\fldZ}{\mathcal{Z}}

\newcommand{\gym}{g\indups{YM}}

\newcommand{\Tr}{\mathop{\mathrm{Tr}}}

\newcommand{\order}[1]{\mathcal{O}(#1)}

\ifx\genfrac\sdflkaj
\newcommand{\atopfrac}[2]{{{#1}\above0pt{#2}}}
\else
\newcommand{\atopfrac}[2]{\genfrac{}{}{0pt}{}{#1}{#2}}
\fi
\newcommand{\sfrac}[2]{{\textstyle\frac{#1}{#2}}}
\newcommand{\half}{\sfrac{1}{2}}
\newcommand{\quarter}{\sfrac{1}{4}}

\newcommand{\indup}[1]{_{\mathrm{#1}}}
\newcommand{\indups}[1]{_{\mathrm{\scriptscriptstyle #1}}}

\newcommand{\matr}[2]{\left(\begin{array}{#1}#2\end{array}\right)}

\newcommand{\lrbrk}[1]{\left(#1\right)}
\newcommand{\bigbrk}[1]{\bigl(#1\bigr)}
\newcommand{\brk}[1]{(#1)}

\newcommand{\comm}[2]{[#1,#2]}

\newcommand{\PTerm}[1]{\{#1\}}
\newcommand{\set}[1]{\{#1\}}
\newcommand{\state}[1]{\mathopen{|}#1\mathclose{\rangle}}

\newcommand{\algr}[1]{\mathfrak{#1}}
\newcommand{\grp}[1]{\mathrm{#1}}

\newcommand{\grU}{\grp{U}}

\newcommand{\alSU}{\algr{su}}

\newcommand{\nn}{\nonumber}
\newcommand{\nln}{\nonumber\\}
\newcommand{\nl}[1][0pt]{\nonumber\\[#1]&\hspace{-4\arraycolsep}&\mathord{}}

\newcommand{\earel}[1]{\mathrel{}&\hspace{-2\arraycolsep}#1\hspace{-2\arraycolsep}&\mathrel{}}
\newcommand{\eq}{\earel{=}}

\def\[{\begin{equation}}
\def\]{\end{equation}}
\def\<{\begin{eqnarray}}
\def\>{\end{eqnarray}}

\newcommand{\hypref}[2]{\ifx\href\asklfhas #2\else\href{#1}{#2}\fi}

\ifx\href\asklfhas\newcommand{\href}[2]{#2}\fi
\newcommand{\arxivno}[1]{\href{http://arxiv.org/abs/#1}{#1}}

\begin{document}

\ifarxiv

\setcounter{page}{0}\thispagestyle{empty}
\begin{flushright}\footnotesize
\texttt{\arxivno{hep-th/0409054}}\\
\texttt{AEI 2004-070}
\end{flushright}
\vspace{2cm}

\begin{center}
{\Large\textbf{\mathversion{bold}Spin Chain for Quantum Strings}\par} \vspace{2cm}

\textsc{Niklas Beisert} \vspace{5mm}

\textit{Max-Planck-Institut f\"ur Gravitationsphysik\\
Albert-Einstein-Institut\\
Am M\"uhlenberg 1, 14476 Potsdam, Germany} \vspace{3mm}

\texttt{nbeisert@aei.mpg.de}\par\vspace*{2cm}\vspace*{\fill}

\textbf{Abstract}\vspace{7mm}

\begin{minipage}{12.7cm}\small
We review and compare the integrable structures in 
$\superN=4$ gauge theory and string theory
on $AdS_5\times S^5$. 
Recently, Bethe ans\"atze for gauge theory/weak coupling 
and string theory/strong coupling were proposed to 
describe scaling dimensions in the $\alSU(2)$ subsector.
Here we investigate the Bethe equations for quantum string theory, 
naively extrapolated to weak coupling. 
Excitingly, we find a spin chain Hamiltonian similar, but not equal, 
to the gauge theory dilatation operator.
\end{minipage}\vspace*{\fill}

\end{center}

\newpage

\else

\DOIsuffix{theDOIsuffix}
\Volume{51}
\Issue{1}
\Month{01}
\Year{2003}
\pagespan{3}{}
\Receiveddate{15 November 2003}
\Reviseddate{30 November 2003}
\Accepteddate{2 December 2003}
\Dateposted{3 December 2003}
\keywords{AdS/CFT correspondence, integrable spin chains, Bethe ansatz.}
\subjclass[pacs]{11.25.Tq,02.30.Ik,75.10.Pq}


\title{Spin Chain for Quantum Strings}

\author[N.~Beisert]{Niklas Beisert%
\footnote{Corresponding author: 
e-mail: \textsf{nbeisert@aei.mpg.de}, 
Phone: +49-331-567-7257,
Fax: +49-331-567-7297}} 
\address{Max-Planck-Institut f\"ur Gravitationsphysik, Albert-Einstein-Institut\\
Am M\"uhlenberg 1, 14476 Potsdam, Germany}

\begin{abstract}
We review and compare the integrable structures in 
$\superN=4$ gauge theory and string theory
on $AdS_5\times S^5$. 
Recently, Bethe ans\"atze for gauge theory/weak coupling 
and string theory/strong coupling were proposed to 
describe scaling dimensions in the $\alSU(2)$ subsector.
Here we investigate the Bethe equations for quantum string theory, 
naively extrapolated to weak coupling. 
Excitingly, we find a spin chain Hamiltonian similar, but not equal, 
to the gauge theory dilatation operator.
\end{abstract}
\maketitle

\fi

\section{Introduction}


When computing scaling dimensions of local operators 
in $\grU(N)$ $\superN=4$ gauge theory one can, 
for convenience, restrict to a number of subsectors. 
The smallest, nontrivial one is the $\alSU(2)$ subsector
with only two scalar fields $\fldZ$ and $\phi$.
A local operator in field theory can now be interpreted as a
state of a spin-$\half$ $\alSU(2)$ spin chain, e.g.
\[
\Tr \fldZ\fldZ\phi\fldZ\fldZ\fldZ\phi\phi\fldZ=
\state{\uparrow\uparrow\downarrow\uparrow\uparrow\uparrow\downarrow\downarrow\uparrow}
\]
In this picture, the planar dilatation operator $\dil$, 
which measures gauge theory scaling dimensions,
maps to the spin chain Hamiltonian $\ham$
\[\label{eq:Dil}
\dil(g)=\len+g^2\ham(g)+\order{1/N},\qquad
g^2=\frac{\gym^2 N}{8\pi^2}\,,
\]
where $\len$ counts the number of spin chain sites.
At leading order, $g=0$, Minahan and Zarembo have shown that 
$\ham$ (alias the one-loop dilatation operator) is the
Hamiltonian of the Heisenberg spin chain
\cite{Minahan:2002ve}.
This Hamiltonian is integrable, i.e.~it is part of a
tower of local charges $\charge_r$, $\ham=\charge_2$,
which commute with the 
$\alSU(2)$ generators $\gen{J}$ 
and with each other (at $g=0$)
\[\label{eq:Int}
\comm{\gen{J}}{\charge_r(g)}=0,\qquad
\comm{\charge_r(g)}{\charge_s(g)}=0.
\]
In \cite{Beisert:2003tq} it was subsequently shown that integrability
extends to next-to-leading order in $g$ (two-loops) and conjectured
that \eqref{eq:Int} might hold exactly in perturbation theory or even beyond.
Based on three basic assumptions \cite{Beisert:2003tq,Beisert:2003jb}
\begin{list}{($iii$)}{\topsep3pt\itemsep0pt\parsep0pt\leftmargin2.5em\labelwidth2.5em}
\item[($i$)] integrability,
\item[($ii$)] proper scaling in the thermodynamic limit and
\item[($iii$)] constraints from Feynman diagrams,
\end{list}
it was possible to construct a \emph{unique} spin chain 
Hamiltonian up to at least fourth order in $g^2$ (five-loops)
\cite{Beisert:2004hm}.%
\footnote{For a review on the dilatation operator, its construction and integrability,
see \cite{Beisert:2004ry}.}
This proposal has been confirmed at three-loops by several independent 
methods: An algebraic construction in a larger subsector
\cite{Beisert:2003ys}, a direct computation in QCD
\cite{Moch:2004pa} which can apparently be lifted to $\superN=4$ SYM
\cite{Kotikov:2004er} and a two-loop computation which 
can be lifted to three-loops by means of multiplet splitting
\cite{Eden:2004ua}.%
\footnote{Also a direct computation in 
a matrix quantum mechanics closely related to $\superN=4$ 
yields the same result \cite{Klose:2003qc}.}

The physical interest in gauge theory scaling dimensions
lies in the AdS/CFT correspondence which relates them to 
energies of string configurations on $AdS_5\times S^5$.
The availability of three-loop gauge theory results
lead to very precise comparisons within 
the (near) BMN and spinning strings proposals
\cite{Berenstein:2002jq,Frolov:2003qc}.
While agreement was demonstrated up to two-loops,
see e.g.~\cite{Beisert:2003xu,Beisert:2003ea,Arutyunov:2003rg}
(c.f.~\cite{Tseytlin:2003ii} for a review),
it soon emerged that there are discrepancies starting at three-loops
\cite{Callan:2003xr,Serban:2004jf}.
In \cite{Beisert:2004hm} these were argued to be due to an order-of-limits problem 
and, when so-called wrapping interactions are taken into account properly, 
agreement might be restored.
Therefore, the AdS/CFT correspondence is not in danger, 
but a direct, perturbative comparison of the sort
proposed in \cite{Berenstein:2002jq,Frolov:2003qc} is invalidated.%
\footnote{Nevertheless, in the strict (i.e.~planar, leading $1/J$)
BMN limit, perturbative gauge and string results do seem to agree.}

Here, various Bethe ans\"atze, for classical as 
well as for quantum models, either enable the comparison 
(spinning strings) or, at least, simplify it drastically (near-BMN).
After reviewing the most up-to-date Bethe ans\"atze, 
for gauge theory as well as for string theory,
we shall consider their weak-coupling regime. 
While for gauge theory the Bethe ansatz is known to be equivalent 
to a spin chain Hamiltonian, we show that the same
is true even for string theory! 
This is remarkable because the string Bethe equations are
sufficiently different from common Bethe equations.
It is possible that the novel spin chain underlying the string Bethe ansatz 
is a suitable description of string theory at weak coupling.

\section{Bethe Ans\"atze}

The Bethe ansatz is a means of finding eigenvalues
$E=Q_2,Q_r$ of the Hamiltonian and commuting charges $\ham=\charge_2,\charge_r$
on an integrable system by solving a set of algebraic equations.
Within the Bethe ansatz, a state is represented by a set
of Bethe roots $\set{u_k}$
which specify the rapidities of the magnon spin-waves making up the state.
The charge eigenvalues $Q_r$ are the sums of the 
contributions $q_r(u_k)$ from the individual spin waves.
In both Bethe ans\"atze, for gauge theory \cite{Beisert:2004hm} 
and string theory \cite{Arutyunov:2004vx}, the charges are given
by the same expressions%
\footnote{It turns out that already ($i$) and ($ii$) together imply
this form of charge eigenvalues $Q_r$.}
\[\label{eq:Charges}
Q_r=\sum_{k=1}^K q_r(u_k),
\qquad
q_r(u_k)=\frac{i}{r-1}\lrbrk{\frac{1}{(x_k^+)^{r-1}}-\frac{1}{(x_k^-)^{r-1}}}.
\]
Here we have defined $x^\pm_k$ as additional representations of the 
Bethe roots $u_k$ via%
\footnote{The map between $x$ and $u$ is a double covering. 
We shall use the branch $x^\pm_k\approx u_k\pm \sfrac{i}{2}$ for $g\approx 0$.}
\[\label{eq:Map}
x_k^\pm=x(u\pm \sfrac{i}{2}),
\qquad
x(u)=\half u+\half u\sqrt{1-2g^2/u^2}\,,
\quad
u(x)=x+\frac{g^2}{2x}\,.
\]
In other words, by specifying one of $u_k,x^+_k,x^-_k$, the others
are defined by \eqref{eq:Map}.

The Bethe equation \cite{Beisert:2004hm} for gauge theory 
is a modification of the one for the Inozemtsev spin chain
\cite{Inozemtsev:2002vb,Serban:2004jf}, it reads%
\footnote{The two products are equivalent upon \eqref{eq:Map}.}
\[\label{eq:BetheGauge}
\lrbrk{\frac{x^+_k}{x^-_k}}^L=
\prod_{\textstyle\atopfrac{j=1}{j\neq k}}^K
\frac{x_k^+-x_j^-}{x_k^--x_j^+}\,
\frac{\displaystyle 1-\frac{g^2}{2x_k^+ x_j^-}}{\displaystyle 1-\frac{g^2}{2x_k^- x_j^+}}
=
\prod_{\textstyle\atopfrac{j=1}{j\neq k}}^K
\frac{u_k-u_j+i}{u_k-u_j-i}\,.
\]
These equations have to be solved subject to the constraints
that there are neither roots at infinity nor coinciding roots.
Furthermore the cyclicity or level matching constraint 
$\charge_1=2\pi m$ with $q_1(u_k)=-i\log x^+_k/x^-_k$ has to be obeyed for
physical states.
Then the $Q_r$ give the eigenvalues of a highest weight state
in a representation $[L-2K]$ of $\alSU(2)$.
The above equations reproduce the spectrum of the gauge theory spin chain model,
which is known up to $\order{g^8}$ (five-loop), 
up to wrapping order $\order{g^{2L-2}}$.

Bethe equations for quantum strings were proposed
in \cite{Arutyunov:2004vx}%
\footnote{The two products are equivalent upon summation over $r$.
Note that the second factor in the first product is inverted as compared to \eqref{eq:BetheGauge}.}
\footnote{It would be important to find a transfer matrix 
from which the Bethe equations follow
as in the case of the gauge equations
\cite{Beisert:2004hm}.}
\<\label{eq:BetheString}
\lrbrk{\frac{x^+_k}{x^-_k}}^L
\eq
\prod_{\textstyle\atopfrac{j=1}{j\neq k}}^K
\frac{x_k^+-x_j^-}{x_k^--x_j^+}\,
\frac{\displaystyle 1-\frac{g^2}{2x_k^- x_j^+}}{\displaystyle 1-\frac{g^2}{2x_k^+ x_j^-}}\,
\lrbrk{\frac{\displaystyle 1-\frac{g^2}{2x_k^- x_j^+}}{\displaystyle 1-\frac{g^2}{2x_k^+ x_j^+}}\,
\frac{\displaystyle 1-\frac{g^2}{2x_k^+ x_j^-}}{\displaystyle 1-\frac{g^2}{2x_k^- x_j^-}}}^{2i(u_k-u_j)}
\\\nn\eq
\prod_{\textstyle\atopfrac{j=1}{j\neq k}}^K
\frac{u_k-u_j+i}{u_k-u_j-i}\,
\exp\lrbrk{ 2i\sum_{r=2}^\infty \bigbrk{\half g^2}^r
\bigbrk{q_{r}(u_k)\,q_{r+1}(u_j)-q_{r+1}(u_k)\,q_{r}(u_j)}}.
\>
These equations were designed to match with the 
equations for the classical string sigma model
\cite{Kazakov:2004qf} in the thermodynamic limit.
They also give correct predictions for 
dimensions of strings in near plane-waves
\cite{Callan:2003xr,Callan:2004uv,Callan:2004ev,McLoughlin:2004dh}.
More excitingly, they reproduce a generic $\sqrt{g}$-scaling
for dimensions at strong coupling. 
As far as we know, these equations yield correct string results for 
large $g$ and $L$ \cite{Arutyunov:2004vx}, nevertheless they are also 
reasonable equations in the weak coupling regime for $g$ and $L$ small.
In fact, it is easy to see that \eqref{eq:BetheString} agrees
with \eqref{eq:BetheGauge} at $\order{g^2}$ (two-loops).
In the remainder of this text we will investigate the 
perturbative regime of both Bethe ans\"atze. 
For simplicity of notation,
we shall distinguish between the gauge 
\eqref{eq:BetheGauge} and string equations \eqref{eq:BetheString}, 
although there is no indication that the string equations 
reproduce string theory results for small $g$.

\section{Spin Chains}

The string equations \eqref{eq:BetheString} are
substantially different from common Bethe equations, 
but at least their abstract form remains the same:
They consist of single particle propagation (l.h.s.) 
and two-particle scattering (r.h.s.) terms which are 
functionally more complicated than usual. 
Furthermore, the equations agree with gauge theory 
(and the corresponding spin chain) up to $\order{g^2}$. 
All this suggests that
there might also be a spin chain formulation for string theory.
Now it was shown that conditions ($i$-$iii$) yield a unique answer
up to and including at least $\order{g^8}$ 
and it does not agree with string theory.
Therefore we need to relax one of the conditions. 
Clearly we cannot modify conditions ($i$) or ($ii$) because the string equations
manifestly have these properties. However there is no reason 
to rely on condition ($iii$) for string theory. This condition
splits up into two statements ($iii$a) and ($iii$b),
see \cite{Beisert:2003tq,Beisert:2003jb,Beisert:2004ry} 
for a detailed discussion of ($i$-$iii$). 
Condition ($iii$a) limits the range of the interaction 
at $\order{g^{2\ell-2}}$ to $\ell+1$ neighbouring spin sites, 
whereas ($iii$b) limits the number of adjacent permutations to $\ell$.

Let us now drop condition ($iii$b) and compute the most general 
Hamiltonian satisfying ($i$,$ii$,$iii$a). In the notation 
using adjacent permutations, see \cite{Beisert:2003tq,Beisert:2004ry}, 
\[\label{eq:PermSym}
\{p_1,p_2,\ldots\}=
\sum_{p=1}^L \opperm_{p+p_1,p+p_1+1}
\opperm_{p+p_2,p+p_2+1}\ldots
\]
we find the three-loop Hamiltonian
\ifarxiv 
(see Tab.~\ref{tab:FiveLoop} for the five-loop contribution,
Tab.~\ref{tab:NonOverlap} for an alternative representation 
and Tab.~\ref{tab:Q3FourLoop} for the four-loop third charge)
\fi
\<\label{eq:Ham}
\ham(g)\eq\sum_{\ell=1}^\infty g^{2\ell-2}\ham_{2\ell-2},
\nln
\ham_0\eq \PTerm{}-\PTerm{1},
\nln
\ham_2\eq 
-2\PTerm{}
+3\PTerm{1}
-\half\bigbrk{\PTerm{1,2}+\PTerm{2,1}},
\nln
\ham_4\eq 
+\brk{\sfrac{15}{2}-\half c_4}\PTerm{}
+\brk{-13+\sfrac{3}{2}c_4}\PTerm{1}
+\brk{\half-\sfrac{3}{4}c_4}\PTerm{1,3}
+\brk{3-\half c_4}\bigbrk{\PTerm{1,2}+\PTerm{2,1}}
\nl
+\half c_4\bigbrk{\PTerm{1,3,2}+\PTerm{2,1,3}}
-\half\bigbrk{\PTerm{1,2,3}+\PTerm{3,2,1}}
-\sfrac{1}{4} c_4\PTerm{2,1,3,2}.
\>
Note that the interaction $\PTerm{2,1,3,2}$ in $\ham_4$ is composed from 
four adjacent permutations whereas ($iii$b) would allow for only three.
This forces $c_4=0$ for gauge theory.

We can now try to adjust the free parameter $c_4$ 
to match the spectrum of $\ham(g)$ with the string Bethe equations. 
Remarkably, this appears to be possible and we find $c_4=1$!
E.g.~for the state with $L=4,K=2$ the Hamiltonian yields
\[D=4+\frac{3\gym^2N}{4\pi^2}-\frac{3\gym^4N^2}{16\pi^4}
+\frac{(42-9c_4)\gym^6N^3}{512\pi^6}\,,
\]
while the string Bethe equations predict $33=42-9$ for the last coefficient. 
To give further support to this observation, we have extended the
analysis to four and five loops.%
\footnote{Investigating the form of the five-loop Hamiltonian and charges,
we notice that the number of adjacent permutations for $\charge_r$ 
at $\order{g^{2\ell-2}}$ is limited to $r+2\ell-4$ 
(plus an additional one at $\ell=1$)
as opposed to $r+\ell-2$ in ($iii$b).}
There we find one and two free parameters, $c_6$ and $c\indup{8a,b}$, which 
can again be adjusted to agree with the predictions from 
\eqref{eq:BetheString}, i.e.~$c_6=c\indup{8a,b}=0$.
For the comparisons we have used an extensive list of two-excitation states,
see Tab.~\ref{tab:TwoEx} for their energy formula.
To check our results we have used one 
paired three-excitation state at $L=7$ and found agreement
\[
D^\pm=7+\frac{5\gym^2N}{8\pi^2}-\frac{15\gym^4N^2}{128\pi^4}
+\frac{95\gym^6N^3}{2048\pi^6}-\frac{155\gym^8N^4}{6144\pi^8}
+\frac{16335\gym^{10}N^5}{1048576\pi^{10}}\,.
\]
\begin{table}\centering
$\displaystyle
D^J_n=J+2
+\sum_{\ell=1}^\infty 
\lrbrk{\frac{\gym^2 N}{\pi^2}\sin^2 \frac{\pi n}{J+1}}^{\ell}
\lrbrk{\frac{(-1)^{\ell-1}(2\ell-2)!}{4^{\ell-1}\ell!(\ell-1)!}+\sum_{k,l=1}^{\ell-1}c_{\ell,k,l}\frac{\cos^{2l} \frac{\pi n}{J+1}}{(J+1)^k}},
$\bigskip\par
\raggedright
$c_{2,1,1}=-1$,
\vspace{0.2cm}\par
$c_{3,k,l}=\matr{ll}{+\sfrac{3}{4}-\half c_4&+\sfrac{1}{2}\\[1.5pt]-\sfrac{3}{4}&+\sfrac{5}{2}}$,
\vspace{0.2cm}\par
$c_{4,k,l}=\matr{lll}{
-\sfrac{5}{8}+\sfrac{7}{12}c_4-c_6&-\sfrac{5}{12}+\sfrac{1}{6}c_4+4c_6&-\sfrac{1}{3}\\[1.5pt]
+\sfrac{3}{4}-\sfrac{3}{4}c_4&- \sfrac{7}{4}+3c_4&-\sfrac{7}{2}\\[1.5pt]
-\sfrac{1}{2}&+\sfrac{59}{12}&-\sfrac{49}{6}
}$,
\vspace{0.2cm}\par
$c_{5,k,l}=\mathnormal{}$\scriptsize$\matr{llll}{
+\frac{35}{64}-\sfrac{5}{8}c_4+\sfrac{1}{32}c_4^2+\sfrac{11}{4}c_6+\sfrac{1}{3}c\indup{8a}-\sfrac{1}{2}c\indup{8b}
  &+\frac{35}{96}-\sfrac{11}{48}c_4-7c_6-6c\indup{8b}
  &+\frac{7}{24}-\sfrac{1}{12}c_4+8c\indup{8b}
  &+\frac{1}{4}\\[3pt]
-\frac{45}{64}+\frac{17}{16}c_4-\frac{3}{16}c_4^2-\sfrac{3}{2}c_6
  &+\frac{185}{96}-\frac{47}{12}c_4+\frac{7}{8}c_4^2+17c_6
  &+\frac{131}{48}-\frac{10}{3}c_4-32c_6
  &+\frac{33}{8}\\[3pt]
+\frac{5}{8}-\frac{3}{4}c_4&-\frac{125}{24}+\frac{17}{2}c_4&-\frac{13}{24}-16c_4&+\frac{81}{4}\\[3pt]
-\frac{5}{16}&+\frac{305}{48}&-\frac{1319}{48}&+\frac{243}{8}
}$.
\caption{Two excitation formula up to five loops with $J=L-K=L-2$.
For gauge theory set $c_4=c_6=c\indup{8a}=c\indup{8b}=0$.
For the string Bethe ansatz set $c_4=1$, $c_6=c\indup{8a}=c\indup{8b}=0$.}
\label{tab:TwoEx}
\end{table}

Finally, we have repeated the numerical comparison of \cite{Beisert:2003ea}
between spinning strings and higher-loop spin chains for the above model
with $c_4=1$. For the three-loop dimensions $\delta''_3$ at $J=4,8,12,16$
(to be compared to Table~1 in \cite{Beisert:2003ea}) we find
the values $0.068651$, $0.068938$, $0.106585$, $0.126870$, respectively. 
The dimension extrapolated to $J=\infty$ as described in 
\cite{Beisert:2003ea} is $0.184\ldots$ which is in $2\%$ agreement with
the energy $0.181347$ of spinning stings 
and thus with the Bethe ansatz \eqref{eq:BetheString}.

\section{Discussion}

Here we have presented a perturbative spin chain Hamiltonian that
agrees with the Bethe ansatz for quantum string theory 
extrapolated to weak coupling. It is very similar to 
the gauge theory Hamiltonian, but has a slightly 
extended form of interactions.
In terms of physics the two Hamiltonians are quite different:
The gauge Hamiltonian does not perturbatively reproduce
the near plane-wave and spinning strings results. 
It was argued that agreement may be restored when 
order-of-limits or, more explicitly, wrapping effects
are taken into account \cite{Beisert:2004hm}.
In contradistinction, 
the string Hamiltonian naively agrees with
near plane-waves and spinning strings in perturbation theory!
This is somewhat disappointing, as one might have hoped that
the `strange' string Bethe equations 
would not have yielded a sensible spin chain at weak coupling. 
Their form would thus have had to be altered to give
a spin chain at weak coupling (presumably precisely the gauge theory spin chain). 
At this point such an interpolating Bethe ansatz does not appear
necessary for consistency reasons any longer. 
Instead, it is a logical possibility that string theory 
is described by \eqref{eq:BetheString} at all values of $g$
and in particular by \eqref{eq:Ham} with $c_4=1$ at weak coupling.%
\footnote{It was shown that the spectrum of string states 
is compatible with the spectrum of gauge theory local operators 
when a suitable dimension formula is chosen
\cite{Bianchi:2003wx,Beisert:2003te}.}
This would be disastrous for the AdS/CFT correspondence which 
requires $c_4=0$ to achieve agreement with gauge theory.
A test of this option might be achieved by computing higher $1/J$ 
corrections in near plane-wave string theory%
\footnote{See \cite{Swanson:2004mk} for first steps in this direction.}
and comparing them to \eqref{eq:BetheString}.
However, it is also possible that $c_4=1$ effectively 
sums up all putative wrapping effects 
and thus describes string theory only in the thermodynamic limit.
In this case, the equivalence between 
\eqref{eq:BetheString} and \eqref{eq:Ham} is still a remarkable
mathematical result which clearly deserves further investigation.

Unfortunately, we do not yet know a good condition 
($iii$b') to replace ($iii$b) which would fix the
parameters values $c_4,\ldots$ uniquely.%
\footnote{It would be interesting to find a
Bethe ansatz for arbitrary parameter values, if it exists at all.}
Certainly, we can adjust them to \eqref{eq:BetheString}, 
but this is not very satisfactory when we try to generalise
our result from the $\alSU(2)$ subsector to bigger ones.
For instance, the Hamiltonian \eqref{eq:Ham} is also a restriction 
of the supersymmetric $\alSU(2|3)$ spin chain investigated 
in \cite{Beisert:2003ys} when we set%
\footnote{Note that $\sigma_2=\sfrac{1}{8}c_4$ in \cite{Beisert:2003ys}
is equivalent to $\sigma_2=\sfrac{3}{4}c_4$ in \cite{Beisert:2004ry}.}
\[\label{eq:su23coeffs}
\sigma_1=-\sfrac{3}{2}c_4,\qquad
\sigma_2=\sfrac{1}{8}c_4,\qquad
\sigma_3=\sfrac{1}{8}c_4,\qquad
\sigma_4=0.
\]
Here it would be interesting to compare to the predictions
for fermionic states in near plane-wave string theory 
\cite{Callan:2004ev,McLoughlin:2004dh}.
In the absence of a higher-loop Bethe ansatz,
the virial and coherent methods 
\cite{Callan:2004dt,Kruczenski:2003gt,Kruczenski:2004kw}
would be very helpful.

\ifarxiv

\subsection*{Acknowledgements}

I thank Matthias Staudacher and Arkady Tseytlin for 
discussions.


\bibliography{micro}
\bibliographystyle{nbshort}

\else

\begin{acknowledgement}
I thank Matthias Staudacher and Arkady Tseytlin for 
discussions.
\end{acknowledgement}

\bibliographystyle{nbshortw}
\bibliography{micro}

\fi

\ifarxiv

\begin{table}[e]\centering\tiny
$\begin{array}{rcl}
\ham_0\eq +\PTerm{}-\PTerm{1}
\\[5pt]
\ham_2\eq 
-2\PTerm{}
+3\PTerm{1}
-\half\bigbrk{\PTerm{1,2}+\PTerm{2,1}}
\\[5pt]
\ham_4\eq 
+\bigbrk{+\sfrac{15}{2}-\half c_4}\PTerm{}
\nl[2pt]
+\bigbrk{-13+\sfrac{3}{2}c_4}\PTerm{1}
\nl[2pt]
+\bigbrk{\half-\sfrac{3}{4}c_4}\PTerm{1,3}
\nl[2pt]
+\bigbrk{+3-\half c_4}\bigbrk{\PTerm{1,2}+\PTerm{2,1}}
\nl[2pt]
+\bigbrk{+\half c_4}\bigbrk{\PTerm{1,3,2}+\PTerm{2,1,3}}
\nl[2pt]
+\bigbrk{-\half}\bigbrk{\PTerm{1,2,3}+\PTerm{3,2,1}}
\nl[2pt]
+\bigbrk{-\sfrac{1}{4} c_4}\PTerm{2,1,3,2}
\\[5pt]
\ham_6\eq
+\bigbrk{-35+\sfrac{13}{4}c_4-6c_6} \PTerm{} 
\nl[2pt]
+\bigbrk{+67+4\alpha_6} \PTerm{1} 
\nl[2pt]
+\bigbrk{-\sfrac{21}{4}-2\alpha_6+\sfrac{11}{12}c_4+4c_6} \PTerm{1,3} 
\nl[2pt]
+\bigbrk{-\sfrac{1}{4}+\sfrac{1}{6}c_4-8c_6}\PTerm{1,4} 
\nl[2pt]
+\bigbrk{-\sfrac{151}{8}-4\alpha_6-\sfrac{79}{12}c_4+16c_6} \bigbrk{ \PTerm{1,2} + \PTerm{2,1} }  
\nl[2pt]
+\bigbrk{+2\alpha_6+c_4-12c_6} \bigbrk{ \PTerm{1,3,2} + \PTerm{2,1,3} }  
\nl[2pt]
+\bigbrk{+\sfrac{1}{4} - \sfrac{5}{12}c_4 + 2c_6 } \bigbrk{ \PTerm{1,2,4} + \PTerm{1,3,4} + \PTerm{1,4,3} +\PTerm{2,1,4} }  
\nl[2pt]
+\bigbrk{+6 + 2\alpha_6 + \sfrac{61}{12}c_4 - 10c_6 } \bigbrk{ \PTerm{1,2,3} + \PTerm{3,2,1} }  
\nl[2pt]
+\bigbrk{-\sfrac{3}{4} - 2\alpha_6 - \sfrac{19}{6}c_4 + 14c_6 } \PTerm{2,1,3,2} 
\nl[2pt]
+\bigbrk{+ \sfrac{9}{8} + 2\alpha_6 + \sfrac{47}{8}c_4 - 9c_6 } \bigbrk{ \PTerm{1,3,2,4} + \PTerm{2,1,4,3} }  
\nl[2pt]
+\bigbrk{-\sfrac{1}{2} - \alpha_6 - \sfrac{19}{8}c_4 + 6c_6 } \bigbrk{ \PTerm{1,2,4,3} + \PTerm{1,4,3,2} + \PTerm{2,1,3,4} + \PTerm{3,2,1,4} }  
\nl[2pt]
+\bigbrk{-\sfrac{5}{8}}\bigbrk{ \PTerm{1,2,3,4} + \PTerm{4,3,2,1} } 
\nl[2pt]
+\bigbrk{-\sfrac{5}{24}c_4 - 2c_6 } \bigbrk{ \PTerm{1,3,2,4,3} + \PTerm{2,1,3,2,4} + \PTerm{2,1,4,3,2} + \PTerm{3,2,1,4,3} }  
\nl[2pt]
+\bigbrk{ +\sfrac{1}{24}c_4 + c_6 } \bigbrk{ \PTerm{2,1,3,2,4,3} + \PTerm{3,2,1,4,3,2} }
\\[5pt]
\ham_8\eq
+\bigbrk{+\sfrac{1479}{8}+4\alpha\indup{8d}-\sfrac{39}{2}c_4+\sfrac{3}{16}c_4^2+86c_6+4c\indup{8a}-10c\indup{8b}}
\PTerm{}
\nl[2pt]
+\bigbrk{-\sfrac{1043}{4}-12\alpha_6+4\alpha\indup{8a}-16\alpha\indup{8d}+\sfrac{535}{16}c_4-\sfrac{3}{8}c_4^2-293c_6-12c\indup{8a}+40c\indup{8b}}
\PTerm{1}
\nl[2pt]
+\bigbrk{-19+8\alpha_6-2\alpha\indup{8a}-4\alpha\indup{8b}+8\alpha\indup{8d}-\sfrac{1667}{96}c_4+\sfrac{51}{32}c_4^2+\sfrac{177}{2}c_6+\sfrac{10}{3}c\indup{8a}-38c\indup{8b}}
\PTerm{1,3}
\nl[2pt]
+\bigbrk{+5+2\alpha_6+4\alpha\indup{8b}+4\alpha\indup{8c}+8\alpha\indup{8d}-\sfrac{5}{3}c_4-\sfrac{19}{16}c_4^2+103c_6-\sfrac{8}{3}c\indup{8a}+16c\indup{8b}}
\PTerm{1,4}
\nl[2pt]
+\bigbrk{+\sfrac{1}{8}+\sfrac{1}{16}c_4+\sfrac{1}{8}c_4^2+10c_6+10c\indup{8b}}
\PTerm{1,5}
\nl[2pt]
+\bigbrk{+11\alpha_6-4\alpha\indup{8a}+2\alpha\indup{8c}+6\alpha\indup{8d}+\sfrac{47}{32}c_4+\sfrac{3}{8}c_4^2+81c_6+\sfrac{10}{3}c\indup{8a}-28c\indup{8b}}
\bigbrk{\PTerm{1,2}+\PTerm{2,1}}
\nl[2pt]
+\bigbrk{-\sfrac{1}{4}+\sfrac{5}{12}c_4-\sfrac{9}{16}c_4^2-6c_6+8c\indup{8b}}
\PTerm{1,3,5}
\nl[2pt]
+\bigbrk{+\sfrac{251}{4}-5\alpha_6+2\alpha\indup{8a}-2\alpha\indup{8c}+\sfrac{277}{16}c_4-\sfrac{7}{4}c_4^2-\sfrac{37}{2}c_6-6c\indup{8a}+43c\indup{8b}}
\bigbrk{\PTerm{1,3,2}+\PTerm{2,1,3}}
\nl[2pt]
+\bigbrk{-3-\alpha_6-2\alpha\indup{8c}-4\alpha\indup{8d}+\sfrac{143}{48}c_4+\sfrac{1}{32}c_4^2-\sfrac{49}{2}c_6+\sfrac{8}{3}c\indup{8a}-2c\indup{8b}}
\bigbrk{\PTerm{1,2,4}+\PTerm{1,3,4}+\PTerm{1,4,3}+\PTerm{2,1,4}}
\nl[2pt]
+\bigbrk{-\sfrac{1}{8}+\sfrac{5}{48}c_4-\sfrac{1}{16}c_4^2-\sfrac{11}{2}c_6-2c\indup{8b}}
\bigbrk{\PTerm{1,2,5}+\PTerm{1,4,5}+\PTerm{1,5,4}+\PTerm{2,1,5}}
\nl[2pt]
+\bigbrk{+\sfrac{41}{4}-6\alpha_6+2\alpha\indup{8a}-4\alpha\indup{8c}-2\alpha\indup{8d}-\sfrac{935}{96}c_4+\sfrac{11}{32}c_4^2-\sfrac{15}{2}c_6+\sfrac{8}{3}c\indup{8a}-3c\indup{8b}}
\bigbrk{\PTerm{1,2,3}+\PTerm{3,2,1}}
\nl[2pt]
+\bigbrk{-\sfrac{107}{2}+4\alpha_6-2\alpha\indup{8a}-8\alpha\indup{8d}-\sfrac{171}{32}c_4+\sfrac{35}{32}c_4^2-\sfrac{25}{2}c_6+\sfrac{26}{3}c\indup{8a}-28c\indup{8b}}
\PTerm{2,1,3,2}
\nl[2pt]
+\bigbrk{+\sfrac{1}{4}+\alpha\indup{8b}-\sfrac{19}{32}c_4+\sfrac{3}{8}c_4^2+2c_6-7c\indup{8b}}
\bigbrk{\PTerm{1,3,2,5}+\PTerm{1,3,5,4}+\PTerm{1,4,3,5}+\PTerm{2,1,3,5}}
\nl[2pt]
+\bigbrk{+\sfrac{183}{4}-6\alpha_6+2\alpha\indup{8a}-2\alpha\indup{8b}+2\alpha\indup{8d}-\sfrac{321}{16}c_4+\sfrac{7}{8}c_4^2-\sfrac{19}{2}c_6-2c\indup{8a}+c\indup{8b}}
\bigbrk{\PTerm{1,3,2,4}+\PTerm{2,1,4,3}}
\nl[2pt]
+\bigbrk{-\sfrac{3}{4}-2\alpha\indup{8b}+\sfrac{53}{32}c_4-\sfrac{1}{16}c_4^2-18c_6-6c\indup{8a}+24c\indup{8b}}
\bigbrk{\PTerm{1,2,5,4}+\PTerm{2,1,4,5}}
\nl[2pt]
+\bigbrk{+1+2\alpha\indup{8b}-\sfrac{17}{8}c_4+\sfrac{1}{8}c_4^2+22c_6+6c\indup{8a}-23c\indup{8b}}
\bigbrk{\PTerm{1,2,4,5}+\PTerm{2,1,5,4}}
\nl[2pt]
+\bigbrk{-\sfrac{51}{2}+\sfrac{5}{2}\alpha_6-\alpha\indup{8a}+\alpha\indup{8b}+3\alpha\indup{8c}-\alpha\indup{8d}}
\bigbrk{\PTerm{1,2,4,3}+\PTerm{1,4,3,2}+\PTerm{2,1,3,4}+\PTerm{3,2,1,4}}
\nl[2pt]
+\bigbrk{-\alpha\indup{8b}}\bigbrk{\PTerm{1,2,3,5}+\PTerm{1,3,4,5}+\PTerm{1,5,4,3}+\PTerm{3,2,1,5}}
\nl[2pt]
+\bigbrk{+\sfrac{35}{4}+\alpha_6+2\alpha\indup{8c}+2\alpha\indup{8d}+\sfrac{7}{8}c_4-\sfrac{1}{32}c_4^2-\sfrac{3}{2}c_6-2c\indup{8a}+3c\indup{8b}}
\bigbrk{\PTerm{1,2,3,4}+\PTerm{4,3,2,1}}
\nl[2pt]
+\bigbrk{+\sfrac{1}{8}c_4-\sfrac{3}{16}c_4^2+c_6+4c\indup{8b}}
\bigbrk{\PTerm{1,4,3,5,4}+\PTerm{2,1,3,2,5}}
\nl[2pt]
+\bigbrk{-\sfrac{7}{8}-\alpha_6+2\alpha\indup{8c}-\sfrac{19}{4}c_4-
\sfrac{1}{4}c_4^2+12c_6}
\bigbrk{\PTerm{1,4,3,2,5}+\PTerm{2,1,3,5,4}}
\nl[2pt]
+\bigbrk{+\sfrac{1}{2}+\alpha_6+\sfrac{49}{24}c_4-\sfrac{3}{16}c_4^2+
\sfrac{13}{2}c_6+4c\indup{8a}-14c\indup{8b}}
\bigbrk{\PTerm{1,3,2,5,4}+\PTerm{2,1,4,3,5}}
\nl[2pt]
+\bigbrk{+\sfrac{5}{8}+\sfrac{1}{2}\alpha_6-\alpha\indup{8c}+3\alpha\indup{8d}+\sfrac{333}{64}c_4-\sfrac{11}{32}c_4^2+14c_6-\sfrac{5}{3}c\indup{8a}-2c\indup{8b}}
\bigbrk{\PTerm{1,3,2,4,3}+\PTerm{2,1,3,2,4}+\PTerm{2,1,4,3,2}+\PTerm{3,2,1,4,3}}
\nl[2pt]
+\bigbrk{+\sfrac{1}{4}-2\alpha\indup{8c}+\sfrac{51}{32}c_4+\sfrac{7}{32}c_4^2+7c_6+4c\indup{8a}-17c\indup{8b}}
\bigbrk{\PTerm{1,2,5,4,3}+\PTerm{3,2,1,4,5}}
\nl[2pt]
+\bigbrk{+\sfrac{1}{4}+\sfrac{\alpha_6}{2}+\alpha\indup{8c}+\sfrac{277}{192}c_4-\sfrac{9}{64}c_4^2-\sfrac{39}{4}c_6-4c\indup{8a}+\sfrac{29}{2}c\indup{8b}}
\bigbrk{\PTerm{1,2,4,3,5}+\PTerm{1,3,2,4,5}+\PTerm{2,1,5,4,3}+\PTerm{3,2,1,5,4}}
\nl[2pt]
+\bigbrk{-\sfrac{1}{2}\alpha_6-\alpha\indup{8c}}
\bigbrk{\PTerm{1,2,3,5,4}+\PTerm{1,5,4,3,2}+\PTerm{2,1,3,4,5}+\PTerm{4,3,2,1,5}}
\nl[2pt]
+\bigbrk{-\sfrac{7}{8}}
\bigbrk{\PTerm{1,2,3,4,5}+\PTerm{5,4,3,2,1}}
\nl[2pt]
+
\bigbrk{+\alpha\indup{8d}+\sfrac{47}{192}c_4+\sfrac{1}{8}c_4^2-2c_6-c\indup{8a}+3c\indup{8b}}
\bigbrk{\PTerm{1,4,3,2,5,4}+\PTerm{2,1,3,2,5,4}+\PTerm{2,1,4,3,2,5}+\PTerm{2,1,4,3,5,4}}
\nl[2pt]
+\bigbrk{-2\alpha\indup{8d}-c_4+\sfrac{5}{32}c_4^2-9c_6+\sfrac{4}{3}c\indup{8a}+2c\indup{8b}}
\bigbrk{\PTerm{2,1,3,2,4,3}+\PTerm{3,2,1,4,3,2}}
\nl[2pt]
+\bigbrk{-2\alpha\indup{8d}-\sfrac{3}{8}c_4-4c_6+8c\indup{8b}}
\bigbrk{\PTerm{1,3,2,5,4,3}+\PTerm{3,2,1,4,3,5}}
\nl[2pt]
+\bigbrk{+2\alpha\indup{8d}}
\bigbrk{+\PTerm{1,3,2,4,3,5}+\PTerm{3,2,1,5,4,3}}
\nl[2pt]
+\bigbrk{-\alpha\indup{8d}-\sfrac{31}{64}c_4+c\indup{8a}}
\bigbrk{\PTerm{1,2,4,3,5,4}+\PTerm{2,1,3,2,4,5}+\PTerm{2,1,5,4,3,2}+\PTerm{4,3,2,1,5,4}}
\nl[2pt]
+\bigbrk{+\sfrac{1}{24}c_4-\sfrac{1}{16}c_4^2-4c\indup{8b}}
\PTerm{2,1,4,3,2,5,4}
\nl[2pt]
+\bigbrk{+\sfrac{5}{192}c_4-\sfrac{1}{64}c_4^2+\sfrac{1}{4}c_6-\sfrac{3}{2}c\indup{8b}}
\bigbrk{\PTerm{1,4,3,2,5,4,3}+\PTerm{2,1,3,2,5,4,3}+\PTerm{3,2,1,4,3,2,5}+\PTerm{3,2,1,4,3,5,4}}
\nl[2pt]
+\bigbrk{+\sfrac{11}{192}c_4+\sfrac{1}{64}c_4^2+\sfrac{3}{4}c_6-\sfrac{5}{2}c\indup{8b}}
\bigbrk{\PTerm{1,3,2,4,3,5,4}+\PTerm{2,1,3,2,4,3,5}+\PTerm{3,2,1,5,4,3,2}+\PTerm{4,3,2,1,5,4,3}}
\nl[2pt]
+\bigbrk{-\sfrac{1}{96}c_4+c\indup{8b}}
\bigbrk{\PTerm{2,1,4,3,2,5,4,3}+\PTerm{3,2,1,4,3,2,5,4}}
\nl[2pt]
+\bigbrk{-\sfrac{1}{96}c_4+c\indup{8b}}
\bigbrk{\PTerm{2,1,3,2,4,3,5,4}+\PTerm{4,3,2,1,5,4,3,2}}
\end{array}$

\caption{Interpolating five-loop Hamiltonian. 
For gauge theory set $c_4=c_6=c\indup{8a}=c\indup{8b}=0$.
For the string Bethe ansatz set $c_4=1$, $c_6=c\indup{8a}=c\indup{8b}=0$.
The parameters $\alpha$ do not influence the spectrum.}
\label{tab:FiveLoop}
\end{table}


\begin{table}[e]\centering\tiny
$\begin{array}{rcl}
\ham_{0,12}\eq 1-\opperm_{12}=\half-\half \vec\sigma_1\cdot\vec\sigma_2
\\[5pt]
\ham_{2,123}\eq 
-2\ham_{0,12}
+\half \ham_{0,13}
\\[5pt]
\ham_{4,1234}\eq 
+\sfrac{15}{2}\ham_{0,12}
-3\ham_{0,13}
+\half \ham_{0,14}
\nl[2pt]
+\bigbrk{-\half-\half c_4}\ham_{0,14}\ham_{0,23}
\nl[2pt]
+\bigbrk{+\half+\sfrac{1}{4}c_4}\ham_{0,13}\ham_{0,24}
\nl[2pt]
+\bigbrk{-\quarter c_4}\ham_{0,12}\ham_{0,34}
\\[5pt]
\ham_{6,12345}\eq
-35\ham_{0,12}
+ \sfrac{35}{2} \ham_{0,13}
- 5\ham_{0,14} 
+ \sfrac{5}{8} \ham_{0,15}
\nl[2pt]
+ \bigbrk{+\sfrac{17}{4}- 2\alpha_6- \sfrac{7}{4}c_4 + 12c_6}\ham_{0,14}\ham_{0,23} 
\nl[2pt]
+ \bigbrk{-\sfrac{7}{8} - \alpha_6 - \sfrac{71}{24} c_4+ 4c_6}\ham_{0,15}\ham_{0,23} 
\nl[2pt]
+ \bigbrk{- 5- \sfrac{7}{4}c_4 }\ham_{0,13}\ham_{0,24}
\nl[2pt]
+ \bigbrk{+\sfrac{3}{4} + 2\alpha_6 + \sfrac{16}{3} c_4- 10c_6}\ham_{0,15}\ham_{0,24} 
\nl[2pt]
+ \bigbrk{+\sfrac{7}{8} + \alpha_6+ \sfrac{11}{4} c_4- 6c_6}\ham_{0,13}\ham_{0,25} 
\nl[2pt]
+ \bigbrk{- \sfrac{3}{4} - 2\alpha_6- \sfrac{127}{24} c_4+ 11c_6}\ham_{0,14}\ham_{0,25} 
\nl[2pt]
+ \bigbrk{+\sfrac{3}{4} + 2\alpha_6 + \sfrac{29}{4}c_4 - 12c_6}\ham_{0,12}\ham_{0,34} 
\nl[2pt]
+ \bigbrk{-\sfrac{7}{8} - \alpha_6- \sfrac{71}{24}c_4 + 4c_6}\ham_{0,15}\ham_{0,34} 
\nl[2pt]
+ \bigbrk{- \sfrac{1}{24} c_4+ 2c_6}\ham_{0,12}\ham_{0,35} 
\nl[2pt]
+ \bigbrk{+\sfrac{7}{8} + \alpha_6+ \sfrac{11}{4} c_4- 6c_6}\ham_{0,14}\ham_{0,35} 
\nl[2pt]
+ \bigbrk{- \sfrac{1}{8} c_4- 3c_6}\ham_{0,12}\ham_{0,45} 
\nl[2pt]
+ \bigbrk{- \sfrac{1}{24} c_4+ 2c_6}\ham_{0,13}\ham_{0,45}
\\[5pt]
\ham_{8,123456}\eq
+\sfrac{735}{4} \ham_{0,12}
-105\ham_{0,13} 
+\sfrac{315}{8} \ham_{0,14}
-\sfrac{35}{4} \ham_{0,15}
+\sfrac{7}{8} \ham_{0,16}
\nl[2pt]
+\bigbrk{ - \sfrac{725}{8} + 5\alpha_6 - 2\alpha\indup{8a} - 2\alpha\indup{8c} - 2\alpha\indup{8d} - \sfrac{355}{48}c_4 + \sfrac{39}{32}c_4^2 + \sfrac{35}{2}c_6 + 12c\indup{8a} - 48c\indup{8b} } \ham_{0,14}\ham_{0,23} 
\nl[2pt]
+\bigbrk{ +\sfrac{307}{8} - 2\alpha\indup{8d} + \sfrac{503}{48}c_4 - \sfrac{19}{32}c_4^2 - \sfrac{37}{2}c_6 - 4c\indup{8a} + 24c\indup{8b} } \ham_{0,13} \ham_{0,24} 
\nl[2pt]
+\bigbrk{ +\sfrac{429}{8} - 4\alpha_6 + 2\alpha\indup{8a} + 2\alpha\indup{8b} + 2\alpha\indup{8c} + 2\alpha\indup{8d} - \sfrac{671}{48}c_4 + \sfrac{11}{32}c_4^2 + 14c_6 + \sfrac{8}{3}c\indup{8a} - 14c\indup{8b} } \ham_{0,15} \ham_{0,24} 
\nl[2pt]
+\bigbrk{ - \sfrac{429}{8} + 4\alpha_6 - 2\alpha\indup{8a} - 2\alpha\indup{8b} - 2\alpha\indup{8c} - 4\alpha\indup{8d} + \sfrac{1249}{96}c_4 - \sfrac{7}{32}c_4^2 - \sfrac{45}{2}c_6 - \sfrac{4}{3}c\indup{8a} + 15c\indup{8b} } \ham_{0,14}\ham_{0,25} 
\nl[2pt]
+\bigbrk{ -\sfrac{1}{8} + 2\alpha\indup{8b} + 4\alpha\indup{8c} + 2\alpha\indup{8d} - \sfrac{107}{24}c_4 - \sfrac{3}{16}c_4^2 + \sfrac{23}{2}c_6 - 2c\indup{8a} + 3c\indup{8b} } \ham_{0,16}\ham_{0,25} 
\nl[2pt]
+\bigbrk{ +\sfrac{1}{8} - 2\alpha\indup{8b} - 4\alpha\indup{8c} - 2\alpha\indup{8d} + \sfrac{427}{96}c_4 + \sfrac{3}{16}c_4^2 - \sfrac{23}{2}c_6 + 2c\indup{8a} - 2c\indup{8b} } \ham_{0,15}\ham_{0,26} 
\nl[2pt]
+\bigbrk{ +\sfrac{409}{8} - 5\alpha_6 + 2\alpha\indup{8a} - 4\alpha\indup{8b} - 2\alpha\indup{8c} - \sfrac{1099}{48}c_4 + \sfrac{23}{32}c_4^2 + \sfrac{49}{2}c_6 + 4c\indup{8a} - 20c\indup{8b} } \ham_{0,12}\ham_{0,34} 
\nl[2pt]
+\bigbrk{ - \sfrac{3}{8} - \alpha_6 + 2\alpha\indup{8b} + 2\alpha\indup{8c} + 2\alpha\indup{8d} - \sfrac{71}{12}c_4 - \sfrac{1}{32}c_4^2 + 21c_6 + 2c\indup{8a} - 11c\indup{8b} } \ham_{0,16}\ham_{0,34} 
\nl[2pt]
+\bigbrk{ - \sfrac{3}{4} - 4\alpha\indup{8b} - 4\alpha\indup{8c} - 4\alpha\indup{8d} + \sfrac{49}{8}c_4 + \sfrac{1}{16}c_4^2 - 34c_6 - 4c\indup{8a} + 26c\indup{8b} } \ham_{0,16}\ham_{0,25}\ham_{0,34} 
\nl[2pt]
+\bigbrk{ +\sfrac{3}{4} + 4\alpha\indup{8b} + 4\alpha\indup{8c} + 4\alpha\indup{8d} - \sfrac{49}{8}c_4 - \sfrac{1}{16}c_4^2 + 34c_6 + 4c\indup{8a} - 26c\indup{8b} } \ham_{0,15}\ham_{0,26}\ham_{0,34} 
\nl[2pt]
+\bigbrk{ - \sfrac{309}{16} + \sfrac{5}{2}\alpha_6 - \alpha\indup{8a} - \alpha\indup{8b} - 2\alpha\indup{8c} - \alpha\indup{8d} + \sfrac{51}{4}c_4 - \sfrac{29}{64}c_4^2 - 14c_6 - \sfrac{14}{3}c\indup{8a} + 21c\indup{8b} } \bigbrk{ \ham_{0,15}\ham_{0,23} + \ham_{0,15}\ham_{0,34} } 
\nl[2pt]
+\bigbrk{ +\sfrac{3}{4} + 4\alpha\indup{8b} + 4\alpha\indup{8c} - \sfrac{105}{16}c_4 - \sfrac{1}{8}c_4^2 + 29c_6 + 4c\indup{8a} - 16c\indup{8b} } \ham_{0,16}\ham_{0,24}\ham_{0,35} 
\nl[2pt]
+\bigbrk{ +\sfrac{319}{16} - 2\alpha_6 + \alpha\indup{8a} + \alpha\indup{8b} + \alpha\indup{8c} + 3\alpha\indup{8d} - \sfrac{1463}{192}c_4 + \sfrac{5}{32}c_4^2 + \sfrac{89}{4}c_6 + 2c\indup{8a} - \sfrac{33}{2}c\indup{8b} } \bigbrk{ \ham_{0,13}\ham_{0,25} + \ham_{0,14}\ham_{0,35} } 
\nl[2pt]
+\bigbrk{ -\sfrac{5}{16} + \sfrac{\alpha_6}{2} - 2\alpha\indup{8b} - 2\alpha\indup{8c} + \sfrac{295}{64}c_4 + \sfrac{1}{32}c_4^2 - \sfrac{69}{4}c_6 - 2c\indup{8a} + \sfrac{15}{2}c\indup{8b} } \bigbrk{ \ham_{0,16}\ham_{0,24} + \ham_{0,16}\ham_{0,35} } 
\nl[2pt]
+\bigbrk{ +\sfrac{3}{8} + \alpha_6 - 2\alpha\indup{8b} - 2\alpha\indup{8c} + \sfrac{71}{12}c_4 + \sfrac{1}{32}c_4^2 - 21c_6 - 2c\indup{8a} + 11c\indup{8b} } \ham_{0,14}\ham_{0,36} 
\nl[2pt]
+\bigbrk{ +\sfrac{3}{4} + 4\alpha\indup{8b} + 4\alpha\indup{8c} - \sfrac{13}{2}c_4 - \sfrac{1}{16}c_4^2 + 30c_6 + 4c\indup{8a} - 18c\indup{8b} } \ham_{0,14}\ham_{0,25}\ham_{0,36} 
\nl[2pt]
+\bigbrk{ +\sfrac{5}{16} - \sfrac{1}{2}\alpha_6 + 2\alpha\indup{8b} + 2\alpha\indup{8c} - \sfrac{437}{96}c_4 - \sfrac{1}{64}c_4^2 + 18c_6 + 2c\indup{8a} - 10c\indup{8b} } \bigbrk{ \ham_{0,14}\ham_{0,26} + \ham_{0,15}\ham_{0,36} } 
\nl[2pt]
+\bigbrk{ -\sfrac{3}{4} - 4\alpha\indup{8b} - 4\alpha\indup{8c} + \sfrac{209}{32}c_4 + \sfrac{3}{32}c_4^2 - \sfrac{59}{2}c_6 - 4c\indup{8a} + 17c\indup{8b} } \bigbrk{ \ham_{0,14}\ham_{0,26}\ham_{0,35} + \ham_{0,15}\ham_{0,24}\ham_{0,36} } 
\nl[2pt]
+\bigbrk{ +\sfrac{3}{8} + 2\alpha\indup{8b} + 2\alpha\indup{8c} + 2\alpha\indup{8d} - \sfrac{95}{48}c_4 - \sfrac{3}{4}c_4^2 + 44c_6 + 6c\indup{8b} } \ham_{0,12} \ham_{0,45} 
\nl[2pt]
+\bigbrk{ +\sfrac{3}{4} + 4\alpha\indup{8d} + \sfrac{115}{48}c_4 - 18c_6 - 8c\indup{8a} + 22c\indup{8b} } \ham_{0,16}\ham_{0,23} \ham_{0,45} 
\nl[2pt]
+\bigbrk{ - \sfrac{1}{16} - \sfrac{1}{2}\alpha_6 + \alpha\indup{8b} + 2\alpha\indup{8c} - \sfrac{59}{24}c_4 + \sfrac{7}{64}c_4^2 - 22c_6 - \sfrac{10}{3}c\indup{8a} + 11c\indup{8b} } \bigbrk{ \ham_{0,12}\ham_{0,35} + \ham_{0,13}\ham_{0,45} } 
\nl[2pt]
+\bigbrk{ - \sfrac{7}{16} - \alpha\indup{8c} - 2\alpha\indup{8d} - \sfrac{29}{192}c_4 + \sfrac{5}{32}c_4^2 + \sfrac{23}{4}c_6 + 4c\indup{8a} - \sfrac{21}{2}c\indup{8b} } \bigbrk{ \ham_{0,16}\ham_{0,23} + \ham_{0,16}\ham_{0,45} } 
\nl[2pt]
+\bigbrk{ - \sfrac{3}{8} - \sfrac{7}{16}c_4 + \sfrac{3}{32}c_4^2 + 9c_6 + 2c\indup{8a} - 9c\indup{8b} } \ham_{0,13}\ham_{0,46} 
\nl[2pt]
+\bigbrk{ +\sfrac{3}{4} + \sfrac{11}{12}c_4 - \sfrac{3}{16}c_4^2 - 14c_6 - 4c\indup{8a} + 18c\indup{8b} } \ham_{0,13}\ham_{0,25} \ham_{0,46} 
\nl[2pt]
+\bigbrk{ +\sfrac{7}{16} + \alpha\indup{8c} + \alpha\indup{8d} - \sfrac{1}{3}c_4 - \sfrac{5}{32}c_4^2 - \sfrac{23}{4}c_6 - 3c\indup{8a} + \sfrac{21}{2}c\indup{8b} } \bigbrk{ \ham_{0,13}\ham_{0,26} + \ham_{0,15}\ham_{0,46} } 
\nl[2pt]
+\bigbrk{ -\sfrac{3}{4} - 2\alpha\indup{8d} - \sfrac{161}{96}c_4 + \sfrac{1}{8}c_4^2 + 16c_6 + 6c\indup{8a} - 18c\indup{8b} } \bigbrk{ \ham_{0,13}\ham_{0,26}\ham_{0,45} + \ham_{0,15}\ham_{0,23}\ham_{0,46} } 
\nl[2pt]
+\bigbrk{ -\sfrac{1}{32}c_4 - \sfrac{1}{32}c_4^2 + 3c\indup{8b} } \ham_{0,12} \ham_{0,56} 
\nl[2pt]
+\bigbrk{ -\sfrac{1}{24}c_4 + \sfrac{1}{16}c_4^2 + 4c\indup{8b} } \ham_{0,12}\ham_{0,34}\ham_{0,56} 
\nl[2pt]
+\bigbrk{ +\sfrac{3}{16} + \alpha\indup{8b} + \alpha\indup{8c} - \sfrac{35}{24}c_4 - \sfrac{1}{64}c_4^2 + 2c_6 - 2c\indup{8b} } \bigbrk{ \ham_{0,12}\ham_{0,46} + \ham_{0,13}\ham_{0,56} } 
\nl[2pt]
+\bigbrk{ - \sfrac{3}{16} - \alpha\indup{8b} - \alpha\indup{8c} + \sfrac{67}{48}c_4 + \sfrac{3}{64}c_4^2 - \sfrac{5}{2}c_6 } \bigbrk{ \ham_{0,12}\ham_{0,36} + \ham_{0,14}\ham_{0,56} } 
\nl[2pt]
+\bigbrk{ +\sfrac{3}{8} + 2\alpha\indup{8b} + 2\alpha\indup{8c} - \sfrac{67}{24}c_4 + \sfrac{1}{32}c_4^2 + 7c_6 - 4c\indup{8b} } \bigbrk{ \ham_{0,12}\ham_{0,36}\ham_{0,45} + \ham_{0,14}\ham_{0,23}\ham_{0,56} } 
\nl[2pt]
+\bigbrk{ - \sfrac{3}{8} - 2\alpha\indup{8b} - 2\alpha\indup{8c} + \sfrac{17}{6}c_4 + \sfrac{1}{32}c_4^2 - 7c_6 } \bigbrk{ \ham_{0,12}\ham_{0,35}\ham_{0,46} + \ham_{0,13}\ham_{0,24}\ham_{0,56} }
\end{array}$

\caption{Alternative representation of the five-loop Hamiltonian 
using non-overlapping permutations.
For gauge theory set $c_4=c_6=c\indup{8a}=c\indup{8b}=0$.
For the string Bethe ansatz set $c_4=1$, $c_6=c\indup{8a}=c\indup{8b}=0$.
The parameters $\alpha$ do not influence the spectrum.}
\label{tab:NonOverlap}
\end{table}


\begin{table}[e]\centering\footnotesize
$\begin{array}{rcl}
\charge_{3,0}\eq
+\bigbrk{+\sfrac{i}{2}}\bigbrk{ \PTerm{1,2} - \PTerm{2,1} }
\\[8pt]
\charge_{3,2}\eq
+\bigbrk{-2i}\bigbrk{ \PTerm{1,2} - \PTerm{2,1} } 
\nl[3pt]
+\bigbrk{+ \sfrac{i}{2}} \bigbrk{\PTerm{1,2,3} - \PTerm{3,2,1}}
\\[8pt]
\charge_{3,4}\eq
+ \bigbrk{ +\sfrac{73i}{8} - \sfrac{i}{2}c_4 } \bigbrk{ \PTerm{1,2} - \PTerm{2,1} } 
\nl[3pt]
+ \bigbrk{ -\sfrac{i}{4} + \sfrac{3i}{8}c_4 } \bigbrk{ \PTerm{1,2,4} + \PTerm{1,3,4} - \PTerm{1,4,3} - \PTerm{2,1,4} } 
\nl[3pt]
+ \bigbrk{ -\sfrac{7i}{2} + \sfrac{i}{4}c_4 } \bigbrk{ \PTerm{1,2,3} - \PTerm{3,2,1} } 
\nl[3pt]
+ \bigbrk{ - \sfrac{i}{8} - \sfrac{i}{2}c_4 } \bigbrk{ \PTerm{1,3,2,4} - \PTerm{2,1,4,3} } 
\nl[3pt]
+ \bigbrk{ -\sfrac{i}{8} - \sfrac{i}{4}c_4 } \bigbrk{ \PTerm{1,2,4,3} - \PTerm{1,4,3,2} + \PTerm{2,1,3,4} - \PTerm{3,2,1,4} } 
\nl[3pt]
+ \bigbrk{+\sfrac{5i}{8}}\bigbrk{ \PTerm{1,2,3,4} - \PTerm{4,3,2,1} }
\nl[3pt]
+ \bigbrk{+\sfrac{i}{8}c_4} \bigbrk{ \PTerm{1,3,2,4,3} + \PTerm{2,1,3,2,4} - \PTerm{2,1,4,3,2} - \PTerm{3,2,1,4,3} }
\\[8pt]
\charge_{3,6}\eq
+\bigbrk{ -\sfrac{373i}{8} - i\alpha_6 + \sfrac{31i}{24}c_4 - 8ic_6 } \bigbrk{ \PTerm{1,2} - \PTerm{2,1} } 
\nl[3pt]
+ \bigbrk{ +\sfrac{5i}{2} - \sfrac{41i}{12}c_4 + 5ic_6 } \bigbrk{ \PTerm{1,2,4} + \PTerm{1,3,4} - \PTerm{1,4,3} - \PTerm{2,1,4} } 
\nl[3pt]
+ \bigbrk{+ \sfrac{i}{8} - \sfrac{i}{12}c_4 + 4ic_6 } \bigbrk{ \PTerm{1,2,5} + \PTerm{1,4,5} - \PTerm{1,5,4} - \PTerm{2,1,5} } 
\nl[3pt]
+ \bigbrk{ +\sfrac{45i}{2} + 2i\alpha_6 + 3ic_4 - 6ic_6 } \bigbrk{ \PTerm{1,2,3} - \PTerm{3,2,1} } 
\nl[3pt]
+ \bigbrk{ +\sfrac{i}{48}c_4 - ic_6 } \bigbrk{ \PTerm{1,3,2,5} - \PTerm{1,3,5,4} + \PTerm{1,4,3,5} - \PTerm{2,1,3,5} } 
\nl[3pt]
+ \bigbrk{ +2i + \sfrac{14i}{3}c_4 - 2ic_6 } \bigbrk{ \PTerm{1,3,2,4} - \PTerm{2,1,4,3} } 
\nl[3pt]
+ \bigbrk{ -\sfrac{i}{4} + \sfrac{5i}{12}c_4 - 2ic_6 } \bigbrk{ \PTerm{1,2,4,5} - \PTerm{2,1,5,4} } 
\nl[3pt]
+ \bigbrk{ +\sfrac{23i}{16} - \sfrac{i}{2}\alpha_6 + \sfrac{15i}{16}c_4 + \sfrac{3i}{2}c_6 } \bigbrk{ \PTerm{1,2,4,3} - \PTerm{1,4,3,2} + \PTerm{2,1,3,4} - \PTerm{3,2,1,4} }
\nl[3pt]
+ \bigbrk{ -\sfrac{i}{4} + \sfrac{19i}{48}c_4 - ic_6 } \bigbrk{ \PTerm{1,2,3,5} + \PTerm{1,3,4,5} - \PTerm{1,5,4,3} - \PTerm{3,2,1,5} } 
\nl[3pt]
+ \bigbrk{ -\sfrac{57i}{8} - i\alpha_6 - \sfrac{29i}{12}c_4 + 5ic_6 } \bigbrk{ \PTerm{1,2,3,4} - \PTerm{4,3,2,1} } 
\nl[3pt]
+ \bigbrk{ +\sfrac{i}{2} + i\alpha_6 + \sfrac{21i}{8}c_4 - 6ic_6 } \bigbrk{ \PTerm{1,4,3,2,5} - \PTerm{2,1,3,5,4} } 
\nl[3pt]
+ \bigbrk{ +\sfrac{i}{16} + \sfrac{i}{2}\alpha_6 + \sfrac{i}{16}c_4 - 3ic_6 } \bigbrk{ \PTerm{1,3,2,4,3} + \PTerm{2,1,3,2,4} - \PTerm{2,1,4,3,2} - \PTerm{3,2,1,4,3} } 
\nl[3pt]
+ \bigbrk{ -\sfrac{7i}{16} - \sfrac{i}{2}\alpha_6 - \sfrac{15i}{8}c_4 + \sfrac{3i}{2}c_6 } \bigbrk{ \PTerm{1,2,4,3,5} + \PTerm{1,3,2,4,5} - \PTerm{2,1,5,4,3} - \PTerm{3,2,1,5,4} } 
\nl[3pt]
+ \bigbrk{ +\sfrac{i}{16} + \sfrac{i}{2}\alpha_6 + \sfrac{17i}{16}c_4 - 3ic_6 } \bigbrk{ \PTerm{1,2,3,5,4} - \PTerm{1,5,4,3,2} + \PTerm{2,1,3,4,5} - \PTerm{4,3,2,1,5} } 
\nl[3pt]
+ \bigbrk{+\sfrac{7i}{8}}\bigbrk{ \PTerm{1,2,3,4,5} - \PTerm{5,4,3,2,1} } 
\nl[3pt]
+ \bigbrk{ +\sfrac{i}{24}c_4 + ic_6 } \bigbrk{ \PTerm{1,4,3,2,5,4} - \PTerm{2,1,3,2,5,4} + \PTerm{2,1,4,3,2,5} - \PTerm{2,1,4,3,5,4} } 
\nl[3pt]
+ \bigbrk{+\sfrac{i}{8}c_4 }\bigbrk{ \PTerm{2,1,3,2,4,3} - \PTerm{3,2,1,4,3,2} } 
\nl[3pt]
+ \bigbrk{ +\sfrac{5i}{24}c_4 + 2ic_6 } \bigbrk{ \PTerm{1,3,2,4,3,5} - \PTerm{3,2,1,5,4,3} } 
\nl[3pt]
+ \bigbrk{ +\sfrac{i}{6} c_4 + ic_6 } \bigbrk{ \PTerm{1,2,4,3,5,4} + \PTerm{2,1,3,2,4,5} - \PTerm{2,1,5,4,3,2} - \PTerm{4,3,2,1,5,4} } 
\nl[3pt]
+ \bigbrk{ -\sfrac{i}{48}c_4 - \sfrac{i}{2}c_6 } \bigbrk{ \PTerm{1,4,3,2,5,4,3} - \PTerm{2,1,3,2,5,4,3} + \PTerm{3,2,1,4,3,2,5} - \PTerm{3,2,1,4,3,5,4} } 
\nl[3pt]
+ \bigbrk{ -\sfrac{i}{48}c_4 - \sfrac{i}{2}c_6 } \bigbrk{ \PTerm{1,3,2,4,3,5,4} + \PTerm{2,1,3,2,4,3,5} - \PTerm{3,2,1,5,4,3,2} - \PTerm{4,3,2,1,5,4,3} } 
\end{array}$

\caption{Interpolating four-loop third charge. 
For gauge theory set $c_4=c_6=0$.
For the string Bethe ansatz set $c_4=1$, $c_6=0$.
The parameter $\alpha_6$ does not influence the spectrum.}
\label{tab:Q3FourLoop}
\end{table}


\fi

\end{document}